\documentclass{article}

\usepackage{arxiv}

\usepackage[utf8]{inputenc} 
\usepackage[T1]{fontenc}    
\usepackage{hyperref}       
\usepackage{url}            
\usepackage{booktabs}       
\usepackage{amsmath}
\usepackage{amsfonts}       
\usepackage{nicefrac}       
\usepackage{microtype}      
\usepackage{lipsum}
\usepackage{graphicx}
\usepackage{listings}
\usepackage{xcolor}
\usepackage{algorithm}
\usepackage{algorithmic}

\definecolor{codegreen}{rgb}{0,0.6,0}
\definecolor{codegray}{rgb}{0.5,0.5,0.5}
\definecolor{codepurple}{rgb}{0.58,0,0.82}
\definecolor{backcolour}{rgb}{0.95,0.95,0.92}

\lstdefinestyle{mystyle}{
  backgroundcolor=\color{backcolour}, commentstyle=\color{codegreen},
  keywordstyle=\color{magenta},
  numberstyle=\tiny\color{codegray},
  stringstyle=\color{codepurple},
  basicstyle=\ttfamily\footnotesize,
  breakatwhitespace=false,         
  breaklines=true,                 
  captionpos=b,                    
  keepspaces=true,                 
  numbers=left,                    
  numbersep=5pt,                  
  showspaces=false,                
  showstringspaces=false,
  showtabs=false,                  
  tabsize=2
}
\graphicspath{ {./images/} }

\title{Toward a Comprehensive Simulation Framework for Hypergraphs: A Python-Based Approach}

\author{
 Quoc Chuong Nguyen \\
  Department of Mathematics\\
  University at Buffalo\\
  New York, NY 14260 \\
  \texttt{quocchuo@buffalo.edu}
  \And 
  Trung Kien Le \\ 
  Department of Applied Physics \\
  Stanford University \\
  Stanford, CA 94305 \\ 
  \texttt{letrkien@stanford.edu} \\
}

\begin{document}
\maketitle
\begin{abstract}
Hypergraphs, or generalization of graphs such that edges can contain more than two nodes, have become increasingly prominent in understanding complex network analysis. Unlike graphs, hypergraphs have relatively few supporting platforms, and such dearth presents a barrier to more widespread adaptation of hypergraph computational toolboxes that could enable further research in several areas. Here, we introduce \textbf{HyperRD}, a Python package for hypergraph computation, simulation, and interoperability with other powerful Python packages in graph and hypergraph research. Then, we will introduce two models on hypergraph, the general Schelling's model and the SIR model, and simulate them with HyperRD. 
\end{abstract}


\section{Introduction}
\label{sec:intro}

Hypergraphs, a natural extension of traditional graphs, have emerged as a powerful mathematical tool for modeling complex relationships and structures across various domains \cite{federico2023, bianconi2021, battiston2021}. Unlike graphs, hypergraphs allow for edges to connect more than two vertices, providing a more expressive representation of intricate dependencies. Therefore, hypergraphs have a profound significance in diverse disciplines, including computer science \cite{santoro2023}, biology \cite{klamt2009}, and social networks \cite{federico2022}. The study of hypergraphs in computer science has far-reaching implications for algorithm design, especially in addressing optimization problems where traditional graph models may prove inadequate, as noted in \cite{dis_program}. Hypergraphs also capture complexity in complex interacting networks and provide a glimpse into qualitatively different emergent dynamics \cite{Zhang_2023}. Shifting the focus to the biological realm, hypergraphs find valuable applications in modeling molecular interactions and regulatory networks. Their unique capability to represent interactions among multiple biological entities, such as genes or proteins, contributes to a more nuanced understanding of the intricate relationships governing biological processes, as highlighted in \cite{genomic}. This dual application underscores the interdisciplinary significance of hypergraphs in both computational and biological domains.

The generalization of hypergraph \cite{two_theorem, introreview} provides a more versatile framework for capturing complex interactions that extend beyond binary relationships \cite{alain}. Therefore, the pervasive integration of hypergraphs in modeling real-world systems underscores the need for sophisticated simulation tools to comprehensively examine and understand their dynamic behavior \cite{qionghai}. Simulation, a pivotal instrument in this context, facilitates an in-depth exploration of the evolution of intricate relationships over time. It proves essential for evaluating the influence of diverse factors on the dynamic nature of hypergraph-based systems. By providing a virtual environment conducive to experimentation, simulation tools play an indispensable role in unraveling the complexities inherent in these relationships, offering insights into emergent properties and behaviors that elude analytical approaches.

For this reason, we introduce a novel Python-based platform designed to simulate and analyze hypergraphs, addressing a gap in current computational resources available to researchers and practitioners in mathematics and computer science. Our platform, distinguished by its near-comprehensive range of features, stands out in its ability to represent, manipulate, visualize, and analyze hypergraphs with a level of versatility and depth not commonly found in existing tools \cite{hat, hypernetx, halp, hmetis, phoenix, hyperg, xgi}. A significant feature of our platform is its interoperability, enabling integration with other established platforms and tools. This feature is instrumental in fostering a universal workspace, where users can leverage the strengths of various platforms to conduct more intricate and expansive research on hypergraphs. By bridging these tools, our platform facilitates a more efficient and cohesive workflow and propels the exploration of hypergraphs into new realms, allowing for more complex, interdisciplinary inquiries and applications.

The paper is organized as follows. In Sec. \ref{sec:review}, we start by surveying some recent platforms to simulate hypergraphs and compare their features with our platform. In Sec. \ref{sec:structure}, we introduce the structure of HyperRD and discuss its modules, which are used to analyze and generate random hypergraphs. Sec. \ref{sec:models} proposes some dynamical models on hypergraph and implements them by HyperRD. Finally, we conclude in Sec. \ref{sec:discuss} with a discussion of the results and an outlook on possible future research directions.

\section{Literature Review}
\label{sec:review}

The development of platforms for simulating normal graphs in applied mathematics and computer science has seen substantial progress over the past decades. One of the most significant contributions has been the NetworkX library in Python, which offers extensive tools for the creation, manipulation, and study of the structure, dynamics, and functions of complex networks \cite{networkx}. NetworkX's ease of use and flexibility have made it a popular choice among researchers. Another notable platform is Gephi, an open-source network visualization software that allows for the exploration and understanding of graph structures through sophisticated graphical representations \cite{gephi}. These platforms have been instrumental in advancing the field of graph theory by facilitating complex analyses and visualizations.

The rationale for developing platforms specifically for simulating hypergraphs arises from the limitations of normal graph simulation tools in capturing multi-way relationships. While normal graphs represent binary relationships between entities, hypergraphs are capable of representing more complex, higher-order interactions among multiple entities simultaneously. This capability is crucial in fields such as computational biology, social network analysis, and complex system modeling, where interactions often occur among more than two entities. Therefore, the development of hypergraph-specific platforms is essential to address the nuanced requirements of these advanced applications, providing more accurate and comprehensive tools for researchers and practitioners. Notwithstanding the commendable attributes of some existing platforms such as HAT \cite{hat}, HyperNetX \cite{hypernetx}, halp, \cite{halp}, hMETIS \cite{hmetis}, Phoenix \cite{phoenix}, HyperG \cite{hyperg}, and XGI \cite{xgi}, each of them has their own limitations. These limitations, in turn, accentuate the compelling need for a Python-based simulation tool characterized by heightened comprehensiveness and versatility. The choice of Python as the language of the proposed simulation platform stems from its widespread adoption and versatility in scientific computing,  ensuring accessibility and adaptability for researchers and practitioners across various disciplines. Python's expressive syntax and rich ecosystem of libraries make it an ideal candidate for developing a tool that caters specifically to the nuanced requirements of hypergraph simulation.

Table [\ref{tab:comparison}] shows a more detailed comparison between our platform (HyperRD) with the others.

\begin{table}[!htbp]
    \caption{Comparison of HyperRD to other hypergraph libraries}
    \centering
    \begin{tabular}{lcccccccc}
        \toprule
        Features & HyperRD & HAT & HyperNetX & halp & hMETIS & Phoenix  & HyperG & XGI \\
        \midrule
        Clustering & \checkmark & & \checkmark & \checkmark & & \checkmark & \checkmark & \checkmark \\
        Visualization & \checkmark & \checkmark & \checkmark & & & & \checkmark & \checkmark \\
        Structure analysis & \checkmark & \checkmark & \checkmark & \checkmark & \checkmark & & \checkmark & \checkmark \\
        Random generator & \checkmark & & \checkmark & & & & \checkmark & \checkmark \\
        Multilayer hypergraph & \checkmark & & & & & & & \checkmark \\
        Integrate other platforms & \checkmark & & & & & & & \\
        \bottomrule
    \end{tabular}
    \label{tab:comparison}
\end{table}

\section{Structure}
\label{sec:structure}

\begin{figure}[htb!]
    \centering
    \includegraphics[width=\textwidth]{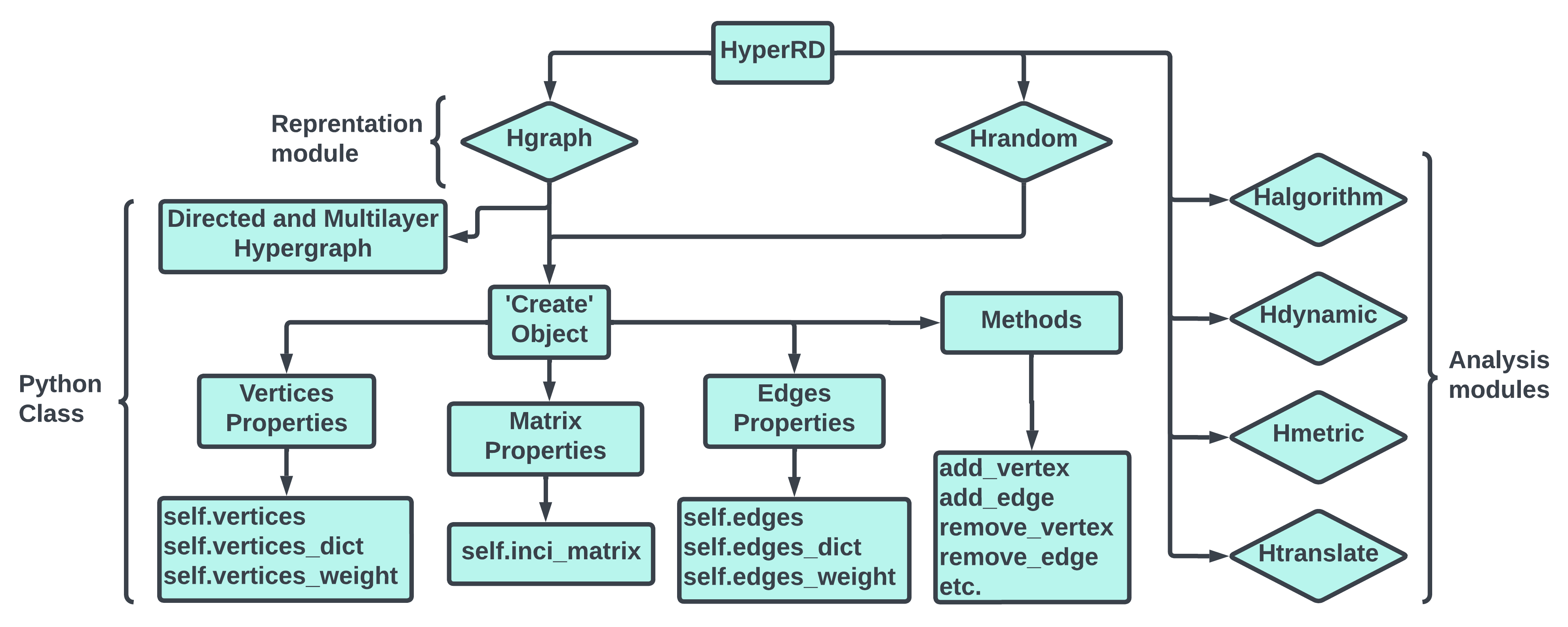}
    \caption{HyperRD consists of two classes of main submodules. Two are for the representation of the hypergraph, and the others are for the analysis of hypergraphs. A hypergraph is constructed through the \texttt{Create} class and its containing features, as illustrated in this figure.}
\label{fig:representation}
\end{figure}

Practically, HyperRD consists of two parts: two submodules to represent the hypergraph, and the others are used for the analysis of the hypergraph as shown in Fig. \ref{fig:representation}.

\subsection{Hypergraph representation}
\label{sub:representation}

In \texttt{Hgraph}, the class \texttt{Create} is designed to encapsulate the properties of an undirected and non-multilayer hypergraph. It contains the following properties:

\begin{itemize}
    \item \textbf{vertices} and \textbf{edges}: Python sets that ensure both vertices and edges are unique and unordered, aligning with the standard mathematical definition of a hypergraph. Sets are efficient for membership testing and eliminating duplicates, key aspects in maintaining the integrity of hypergraph structures \cite{cormen}.
    \item \textbf{vertices\_dict} and \textbf{edges\_dict}: Python dictionaries that provide a mapping mechanism to store additional attributes associated with vertices and edges, respectively. This usage aligns with the need for extending hypergraph representation to store labels,  e.g. indexes of vertices and edges for computation, thereby enhancing the model's applicability to real-world scenarios \cite{claude}.
    \item \textbf{vertices\_weight} and \textbf{edges\_weight}: the implementation of Python lists to store weights for vertices and edges suggesting that the hypergraph is intended to handle weighted scenarios, such as in network flow problems or to represent the strength of connections. Lists provide an ordered and mutable collection of items, which is suitable for scenarios where weights might need to be updated or processed sequentially \cite{jon}.
    \item \textbf{inci\_matrix}: suggest an intention to use an incidence matrix for representation. In hypergraph theory, this matrix is particularly useful for computational analysis of hypergraphs, such as determining vertex-edge connectivity and implementing algorithms for traversal or search within the hypergraph \cite{alain, qionghai}.
\end{itemize}

Random hypergraphs are also important objects studied in hypergraph theory. In essence, random hypergraphs are generalizations of random graphs \cite{bipartite, kuniform}. One of the most renowned models in this domain is the Erdős–Rényi model \cite{erdos}, which extends to hypergraphs by randomly adding hyperedges among a fixed set of vertices. Another significant model is the random $k$-uniform hypergraph, which is a hypergraph where each edge (or hyperedge) connects exactly $k$ vertices. Many studies have shown that random hypergraphs find applications in computational biology and network science, providing insights into the structural properties of complex biological networks \cite{klimm, feng, jose}.

Therefore, owing to the richness of hypergraphs, we present several algorithms for generating random hypergraphs. Throughout the submodule \texttt{Hrandom}, a user can generate hypergraphs via some functions:

\begin{itemize}
    \item \textbf{simple\_matrix}: each element of an incidence matrix will be set to value 1 randomly with probability $p$.
    \item \textbf{simple\_bipartite}: create a random bipartite-featured hypergraph with probability $p$.
    \item \textbf{simple\_powersets}: create a random hypergraph with probability $p$, based on the random choice of all possible edges (the power set of vertices).
    \item \textbf{simple\_order}: similar to \texttt{simple\_powersets}, but the degree of each hyperedge has a bound (the cardinality of the hyperedge is less than or equal to a fixed integer).
    \item \textbf{k\_uniform}: create a $k$-uniform hypergraph with probability $p$.
\end{itemize}

In addition to undirected hypergraph models, \texttt{HyperRD} provides the classes \texttt{Create\_Direct} and \texttt{Multilayer} that inherit from \texttt{Create} and new methods appropriate for each type of hypergraphs. 

\begin{itemize}
    \item \texttt{Create\_Direct}: Construct directed hyperedges for a given hypergraph $H$ from user-specified starting set and ending set. Directed hypergraph models are useful in complex networks with non-reciprocal interactions. 
    \item \texttt{Multilayer}: Construct multilayer hypergraph by adding graphs into a list \textbf{self.layers}, and list of inter-edges \textbf{self.interlink} between layers. Multilayer hypergraph networks are important in modelling multiple types of relationships between components, such as bio-chemical networks involving multiple pathways or logistic networks comprised of multiple modes of transportations. 
\end{itemize}



\subsection{Analysis tools}
\label{sub:analysis}

There are four HyperRD submodules to analyze and visualize hypergraphs, named as \texttt{Halgorithm}, \texttt{Hdynamic}, \texttt{Hmetrix}, \texttt{Htranslate}. The main function of each submodule is:

\begin{itemize}
    \item \texttt{Halgorithm}: provide some basic algorithms on hypergraphs.
    \item \texttt{Hdynamic}: used to study the dynamical systems on hypergraphs.
    \item \texttt{Hmetric}: contains crucial metrics.
    \item \texttt{Htranslate}: transform the \texttt{Create} object to other platforms' object for deeper analysis.
\end{itemize}

\paragraph{Hypergraph metrics} In \texttt{Hmetric}, one can analyze several relevant metrics of hypergraphs such as:

\begin{itemize}
    \item \textbf{Density:} is defined as the number of edges divided by the number of possible edges that a hypergraph can have. That is
    \begin{align}
        \textrm{D(H)} = \frac{\textrm{no. of edges}}{2^{\textrm{no. of vertices}} - 1}
    \end{align}
    \item \textbf{Girth:} the smallest non-empty hyperedge in the hypergraph. For example, if $H$ has edges of size $\{ 2, \cdots, k\}$ then the girth of $H$ is 2. 
    \item \textbf{Average vertex degree:} the vertex degree of a hypergraph is the number of hyperedges connected to the vertex. The average degree is then defined as the sum of all vertices' degrees divided by the length of \textbf{degree\_vertex}. 
    \item \textbf{Average edge size:} similar to the vertex degree, one can assign a notion of average edge size by dividing the total length of all edges in $H$ by the number of edges. 
    \item etc.
\end{itemize}

\paragraph{Hypergraph Algorithms} In HyperRD, the submodule \texttt{Halgorithm} contains algorithms for basic analysis of hypergraph structures. Some of them are listed below:

\begin{itemize}
    \item \textbf{components\_connected}: return a list of connected components in a hypergraph, as shown in Algo. \ref{algo:connected}.
    \item \textbf{simple\_reduction}: reduce an arbitrary hypergraph to a simple form \cite{alain}, as shown in Algo. \ref{algo:reduce}.
    \item \textbf{graph\_expansion}: expand a hypergraph to a normal graph via the clique expansion or star expansion \cite{qionghai}, as shown in Algo. \ref{algo:expand}
    \item etc.
\end{itemize}

\paragraph{Connectivity}

A major feature of HyperRD is its ability to integrate with and take advantage of the rich Python ecosystem for graphs and hypergraphs via the submodule \texttt{Htranslate}. Here, the instance of the \texttt{Create} can be translated into \texttt{HypernetX}'s \textbf{Hypergraph()}, thereby combining the features of \texttt{HyperRD} with these libraries. For example, visualization of hypergraphs can be performed with ease through \texttt{HypernetX} method \textbf{drawing.draw()} after a translate application, which enables quick visualization of a user's instance of \texttt{Create}. At this moment, \texttt{Htranslate} provides the connection to the following platforms: 

\begin{itemize}
    \item \texttt{NetworkX}: a universal Python platform for normal graphs.
    \item \texttt{HypernetX}: used for modularity and clustering on hypergraphs.
    \item \texttt{XGI}: used for dynamics on hypergraphs.
\end{itemize}

\paragraph{Hypergraph dynamics} The higher-order structures in a hypergraph form a fertile ground for novel, emergent dynamics that might not exist in a graph. To provide tools for understanding hypergraph dynamics, HyperRD includes a submodule \texttt{Hdynamic} to simulate random walk on hypergraphs via constructing a probability transition matrix \cite{PhysRevE.101.022308} between nodes $i$ and $j$
\begin{align}
    T_{ij} = \frac{(e\hat{C}e^T)_{ij} - A_{ij}}{\sum_{\ell \neq i} (e\hat{C}e^T)_{i\ell} - k_i^H}
\end{align}
Where $e$ is the incidence matrix, $C = e^T e$ is the hyperedges matrix and $\hat{C}$ is the diagonal matrix of $C$, $A = e e^T$ is the adjacency matrix and $k_i^H = \sum_{\ell \neq i} A_{i\ell} $ is the hyperdegree of $i$-th node. If the random walk is "lazy", one can compute the transition matrix by reducing the numerator to $(e\hat{C}e^T)_{ij}$.



\section{Models}
\label{sec:models}



\subsection{The general Schelling's model of segregation}
\label{sub:schelling_model}

Thomas C. Schelling's pioneering work on segregation models has been a cornerstone in understanding social dynamics through mathematical simulations. In his paper \cite{schelling}, Schelling introduced an agent-based model illustrating how individual preferences, even when not overtly discriminatory, can lead to significant segregation patterns. This model, employing a simple checkerboard to simulate neighborhood dynamics, underscored the unintended consequences of individual choices in societal structures. Further developments by Sakoda, who extended Schelling’s concepts in his paper \cite{sakoda}, and later researchers, incorporated more complex variables and computational techniques, enriching the model’s applicability to diverse social phenomena. Clark and Fossett's review in \cite{clark} and other works \cite{schelling_complexity, schelling_network} provided a comprehensive analysis of the model’s evolution, highlighting its continued relevance in urban studies and policy-making. However, this model does not capture the higher-order relationship between individual actions and collective patterns since agents in one society tend to interact in a group, rather than typical pairwise interaction. To extend this model to higher-order interaction,  we propose the general Schelling's model on a hypergraph as: 

Given the hypergraph \(H = (E, V)\), where \(V = \{v_1, v_2, ..., v_{N}\}\) and \(E = \{e_1, e_2, ..., e_{M}\} \text{, } e_{j} \subset V \text{ }\) for all \(j\). Each node will be assigned one label \(a_{i}\) (via function \(F: V \rightarrow A\)) in the label set \(A = \{a_{1}, a_{2}, ...\}\) such that

\begin{align}
    \bigcup_{i=1}^{|A|} V_{i} \subset V \text{ where } V_{i} = \{v | v \in V \text{ and } F(v) = a_{i} \} \text{ and } V_{i} \cap V_{j} = \emptyset \text{, } \forall i, j
\end{align}

The original version of this model can be easily derived from this definition whereas \(|e_{j}| = 2 \). For the simplification, we only consider A to be finite and there are some unlabeled vertices (i.e., \(\sum_{i = 1}^{|A|} |V_{i}| < |V|\)) and do not allow the one-vertex hyperedge (i.e., \(|e_{j}| > 1\)). 

For each iteration, the process of Schelling's model is described below:

\begin{enumerate}
    \item Select one labeled vertex randomly.
    \item  Calculate the neighborhood coefficient of this vertex via the formula:
        \begin{align}
            G(v) = \frac{1}{K_{v}}\sum_{j=1 \text{, } v \in e_{j}}^{|E|} \frac{|N_{j}(v)|}{|e_{j}| - 1}
        \end{align}

        where,

        \begin{align*}
            & N_{j}(v) = \{v' | v' \neq v \text{, } F(v') = F(v) \text{, and } v' \in e_{j} \} \\
            & K_{v} \text{ is the number of hyperedge that contains v}
    \end{align*}
    \item If \(G(v) < \tau \text{ (} 0 \leq \tau \leq 1)\), assign randomly the label of the chosen vertex to one unlabeled vertex, and the chosen vertex in step 1 will become unlabeled.
    \item Repeat step 1 to 3 until every node has $G(v) \geq \tau$ or after a fixed number of iteration.
\end{enumerate}

\begin{figure}[htb!]
    \includegraphics[width = 0.5\textwidth]{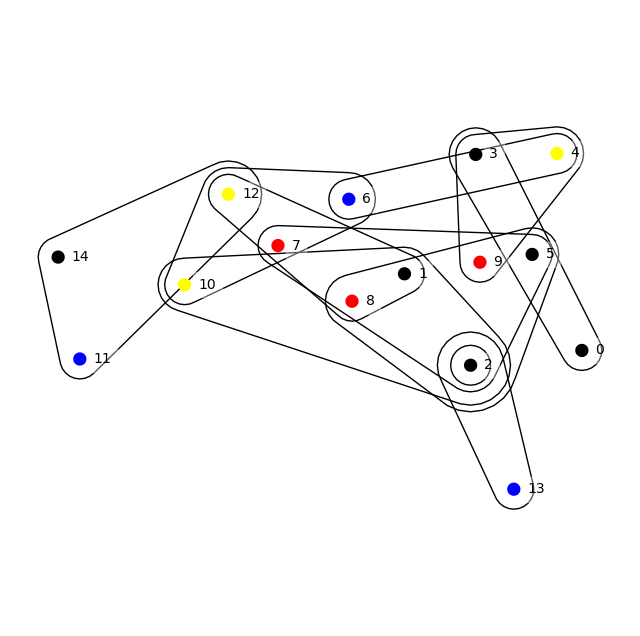}
    \includegraphics[width = 0.5\textwidth]{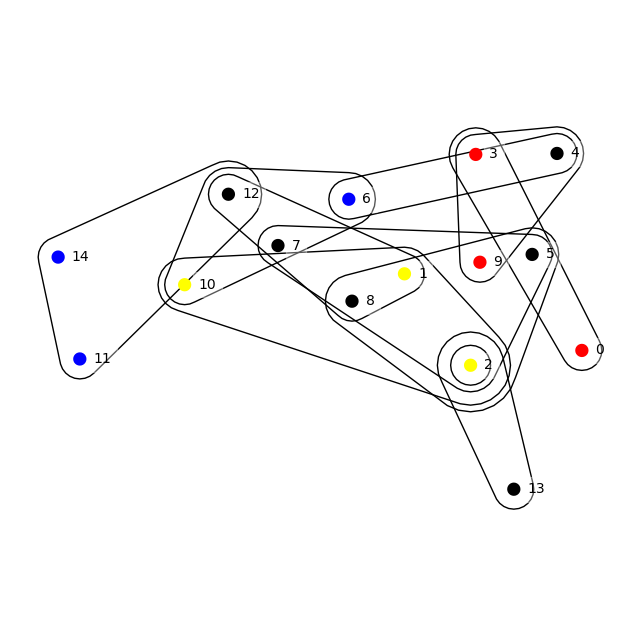}
    \caption{In the implementation of the general Schelling's model, we consider a system consisting of 15 nodes distributed into three distinct labels, with each label assigned to 3 nodes. The initial configuration of these nodes is randomized. We ran the simulation of our model for 100 iterations. The initial state of the node configuration is depicted on the left side of the figure, while the right side illustrates the final state after 100 iterations. It is observed that nodes sharing identical labels tend to aggregate into a single hyperedge, demonstrating the model's inherent clustering dynamics.}
    \label{fig:schelling}
\end{figure}

Observing Fig. \ref{fig:schelling}, the simulation setup comprising 15 nodes, 3 labels, and a randomly initialized state, a clustering phenomenon emerges after 100 iterations. This outcome aligns with the expectations set forth by classical models, yet it introduces a novel characteristic: a convergence of similarly labelled nodes into a single hyperedge. Detailed implementation of this model can be found in the appendix of this paper. \ref{lst:schelling_code}.

\pagebreak 

\subsection{The SIR model on hypergraph}
\label{sub:sir_model}

The traditional Susceptible-Infected-Recovered (SIR) model on normal networks, while a cornerstone in epidemiological modeling, exhibits limitations particularly in representing complex multi-agent interactions, a gap that hypergraph models aim to fill. For instance, the epidemic model on normal networks, as elaborated by Pastor-Satorras et al. \cite{pastor}, explores the spread of infectious diseases in various network topologies, emphasizing the critical role of network structure on disease dynamics. However, these models often overlook scenarios where contagion can occur via group interactions. To address these limitations, researchers have proposed extending the SIR model to hypergraphs. For example, Landry and Restrepo demonstrated the adaptation of the Susceptible-Infected-Susceptible (SIS) model to hypergraphs \cite{landry}, showcasing how it can more accurately reflect the spread of diseases in settings where group interactions are prevalent, such as social gatherings or community clusters. The hypergraph-based SIR model, therefore, not only offers a more realistic framework for simulating disease spread in complex social systems but also provides insights that could be instrumental in shaping effective public health policies and intervention strategies. Here we consider the simple model where the update of the state of each node after one iteration (which is S, I, or R) is given by the probabilities:

\begin{itemize}
    \item \textit{Infection probability}: $P(S \rightarrow I) = 1 - (1 - \beta)^{k_{i}}$  where $k_{i}$ is the number of infected individuals in the same hyperedge, and $\beta$ is the transmission rate.
    \item \textit{Recover probability}: $P(I \rightarrow R) = \gamma $, where $\gamma$ is the recover rate.
\end{itemize}

for all nodes $i$.

\begin{figure}[h!]
    \centering
    \includegraphics[width = 0.45\textwidth]{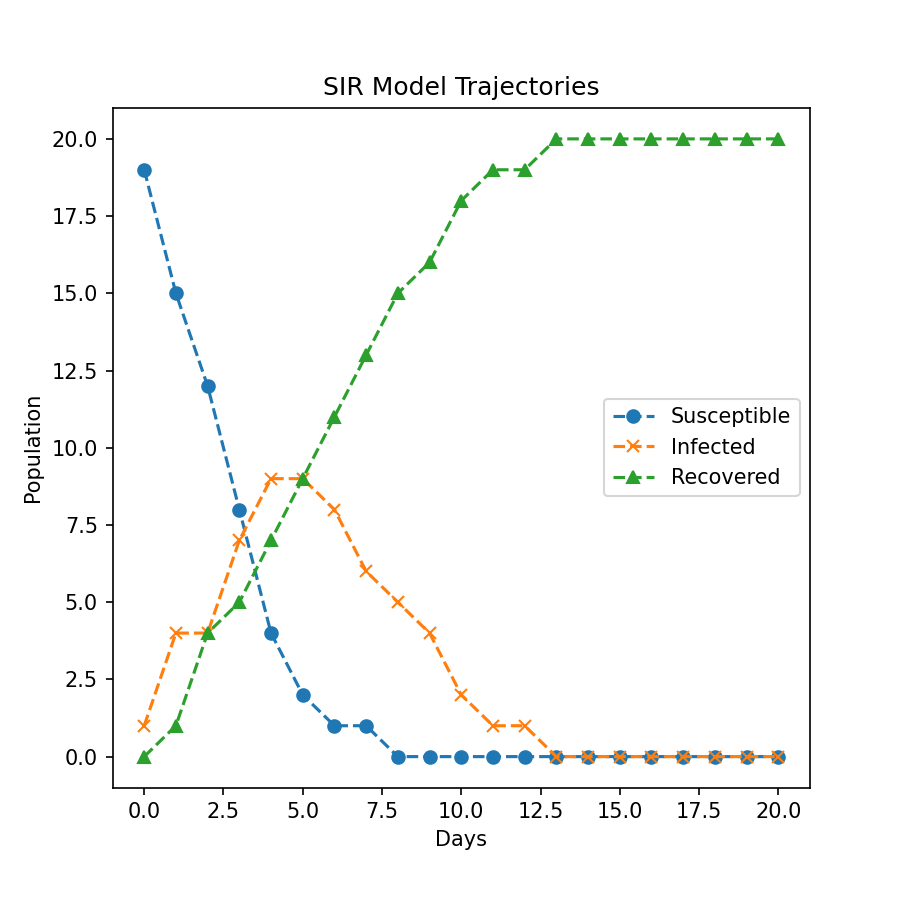}
    \includegraphics[width = 0.45\textwidth]{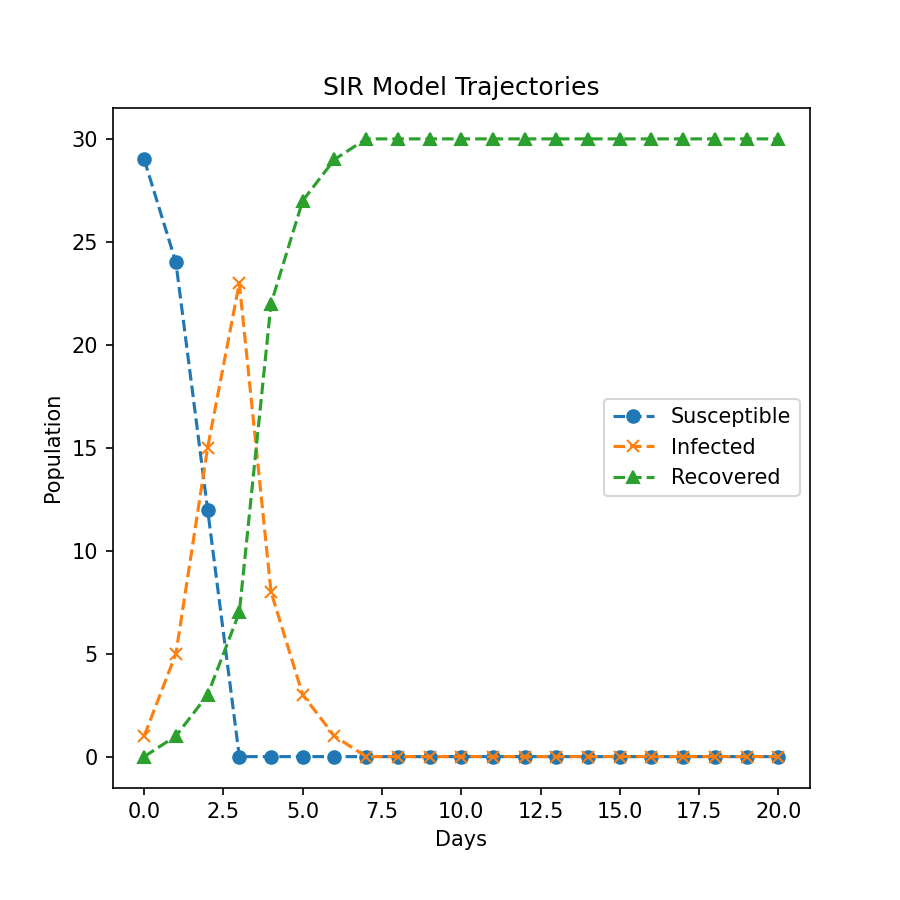}
    \caption{The SIR model on a hypergraph, designed for a maximum interaction order of 3 (3-hyperedges), is depicted for two distinct population sizes: 20 individuals (illustrated on the left) and 30 individuals (shown on the right). In this model, the hyperedges are constrained to a maximum order of 3. The parameters governing the model include a transmission rate $\beta$ set at 0.4 and a recovery rate $\gamma$ at 0.1. The simulation runs for 20 iterations/days.}
    \label{fig:sir}
\end{figure}

Fig. \ref{fig:sir} presents the simulation results of the SIR model on a hypergraph, illustrating two scenarios: one with a population of 20 individuals (shown on the left) and the other with 30 individuals (displayed on the right). In these simulations, the maximum order of the hyperedges is set to 3, with the transmission rate parameter $beta$ at 0.4 and recovery rate parameter $\gamma$ at 0.1, over a total of 20 iterations. For an in-depth understanding of this model's Python implementation, refer to the appendix of this paper.  \ref{lst:sir_code}.

\section{Discussion}
\label{sec:discuss}

In conclusion, HyperRD is the platform designed to simulate hypergraphs, and a cornerstone of this platform's utility lies in its robust connectivity with other platforms and its inherent versatility. This connectivity not only facilitates integration with diverse software ecosystems but also promotes interdisciplinary collaborations, thereby broadening the scope of potential applications.

Moreover, the platform's design philosophy emphasizes continual evolution, ensuring that it remains at state-of-the-art. We will keep regularly upgrading the algorithms and incorporating new features to enable the platform to adapt to the ever-changing landscape of computational needs and user requirements. These include leveraging network data or statistical tools for reconstructing hypergraphs \cite{Young2021}, as well as comprehensive topological \cite{Ghorbanchian2021, Barbarossa2020} and spectral analysis \cite{Banerjee2021} capabilities for fast and powerful structure analysis of large hypergraph networks. Integration of our platform with existing Python frameworks such as \texttt{tensorflow}, \texttt{pytorch} or \texttt{CVXPY} is also possible, furthering our platform's capabilities for interdisciplinary approaches. 

Finally, the exploration of new applications remains a pivotal aspect of this platform. By actively seeking novel use cases, particularly in areas that have yet to fully leverage the power of hypergraph analysis, this platform can significantly contribute to a variety of fields. Some of its applications we will focus on after this paper are synchronization dynamics \cite{Zhang_2023}, directed and non-reciprocal interactions \cite{Gallo2022}, and multilayer dynamics \cite{Alaluusua2023}. Furthermore, applications such as optimizing hypergraph structures for pre-determined goals via tailored loss functions - as already done for graphs to design quantum experiments \cite{RuizGonzalez2023digitaldiscoveryof} - are also viable. 








\section{Acknowledgments}
\label{sec: acknowledgments}

Q.C. and T.K. coordinated the project. Q.C. is the leading developer. All authors contributed to the library and wrote the article.

\section{Data availability statement}
The source code that supports the findings of this study is openly available at the following URL/DOI: 
\begin{center}
  \url{https://github.com/ChuongQuoc1413017/Hypergraph_RD}
\end{center}

\bibliographystyle{unsrt}  
\bibliography{references}

\section{Appendix}
\label{sec:appendix}

\subsection{Python codes}
\label{sub:code}

This is the code for the Schelling's model above. \ref{sub:schelling_model}

\begin{lstlisting}[language=Python, caption=The code for the Schelling's model., label= lst:schelling_code]
from HyperRD.Hrandom import *
from HyperRD.Halgorithm import *
from HyperRD.Hgraph import *
from HyperRD.Hdynamic import *
from HyperRD.Htranslate import *
import matplotlib.pyplot as plt

class Schelling_Model(object):
    # init function to declare class variables
    def __init__(self, n_red, n_blue, n_yellow, graph):
        """Random system initialisation.
        RED    =  0
        BLUE   =  1
        YELLOW =  2
        EMPTY  = -1
        """
        array = np.zeros(len(graph.vertices))
        array[n_red: n_red + n_blue] = 1
        array[n_red + n_blue: n_red + n_blue + n_yellow] = 2
        array[n_red + n_blue + n_yellow:] = -1
        self.graph = graph
        self.array = np.array(array)
        np.random.shuffle(self.array)
    
    def running(self, iteration, tol):
        '''Running the model'''
        total = len(self.array)
        for i in range(iteration):
            citizen = np.where(self.array >= 0)[0]
            start = np.random.choice(citizen, 1, replace = False)[0]
            if self.neighbor_coefficient(start) < tol:
                empty = np.where(self.array < 0)[0]
                end = np.random.choice(empty, 1)[0]
                if self.neighbor_coefficient(end) != 2:
                    self.array[start], self.array[end] = self.array[end], self.array[start]
    
    def neighbor_coefficient(self, start):
        '''calculate the neighbor coefficient'''
        edges = self.graph.edges
        coeff = np.array([])
        for edge in edges:
            if start in edge and len(edge) > 1:
                egde_coeff = self.array[list(edge)]
                neighbor = egde_coeff == self.array[start]
                coeff = np.append(coeff, (np.sum(neighbor) - 1) / (len(edge) - 1))
        return np.mean(coeff) if len(coeff) > 0 else 2

# Initialize hypergraph
graph = simple_order(15, 3, 0.02)
graph.edges

# Initialize model
model = Schelling_Model(3, 3, 3, graph)

# Run the simulation
model.running(100, 0.2)

# Visualize
colors = []
for i in model.array:
    if i == 0:
        colors.append('red')
    elif i == 1:
        colors.append('blue')
    elif i == 2:
        colors.append('yellow')
    else:
        colors.append('black')
graph_x = hyperrd_to_hypernetx(model.graph)
hnx.drawing.draw(graph_x, 
                 with_edge_labels = False, 
                 layout_kwargs = {'seed': 39}, edges_kwargs={
     'edgecolors': 'black'
    },
                 nodes_kwargs={
     'facecolors': colors
    })
\end{lstlisting}

This is the code for the SIR model above. \ref{sub:sir_model}

\begin{lstlisting}[language=Python, caption=The code for the SIR model., label=lst:sir_code]
from HyperRD.Hrandom import *
from HyperRD.Halgorithm import *
from HyperRD.Hgraph import *
from HyperRD.Hdynamic import *
from HyperRD.Htranslate import *
import random
import numpy as np
import matplotlib.pyplot as plt

def compute_mutual_information(trajectory1, trajectory2, bins=10):
    # Compute histograms
    hist_1, _ = np.histogram(trajectory1, bins=bins, density=True)
    hist_2, _ = np.histogram(trajectory2, bins=bins, density=True)
    
    # Compute joint histogram
    joint_hist, _, _ = np.histogram2d(trajectory1, trajectory2, bins=bins, density=True)
    
    # Compute mutual information
    eps = np.finfo(float).eps
    mutual_info = np.sum(joint_hist * np.log2((joint_hist + eps) / (np.outer(hist_1, hist_2) + eps)))
    
    return mutual_info

# Parameters
num_vertices = 20
maximum_hyperedge_size = 3
infection_probability = 0.4
recovery_probability = 0.1
num_steps = 20

# Initialize hypergraph
vertices = range(num_vertices)
graph = simple_order(num_vertices, maximum_hyperedge_size, 0.02)

# Initialize states
states = ['S'] * num_vertices  # S: Susceptible, I: Infected, R: Recovered
initial_infected = random.sample(vertices, 1)
for i in initial_infected:
    states[i] = 'I'
    
# Lists to store the trajectory
susceptible_traj = [len([v for v in states if v == 'S'])]
infected_traj = [len([v for v in states if v == 'I'])]
recovered_traj = [len([v for v in states if v == 'R'])]

# Simulation function
def simulate_step():
    new_states = states.copy()
    for e in graph.edges:
        infected = [v for v in e if states[v] == 'I']
        if infected:
            for v in e:
                if states[v] == 'S' and random.random() < 1 - (1 - infection_probability)**len(infected):
                    new_states[v] = 'I'
                elif states[v] == 'I' and random.random() < recovery_probability:
                    new_states[v] = 'R'
    return new_states

# Run the simulation
for step in range(num_steps):
    states = simulate_step()
    susceptible_traj.append(len([v for v in states if v == 'S']))
    infected_traj.append(len([v for v in states if v == 'I']))
    recovered_traj.append(len([v for v in states if v == 'R']))

# Plot the trajectories
plt.plot(susceptible_traj, label='Susceptible')
plt.plot(infected_traj, label='Infected')
plt.plot(recovered_traj, label='Recovered')
plt.title('SIR Model Trajectories')
plt.xlabel('Days')
plt.ylabel('Population')
plt.legend()
plt.show()

# Print the computed mutual information
mutual_info = compute_mutual_information(susceptible_traj, infected_traj)
print(f'Mutual Information between Susceptible and Infected: {mutual_info}')
\end{lstlisting}

\subsection{Algorithms}
\label{sub:algo}

\begin{algorithm}[htb!]
\caption{Find Connected Components in a Hypergraph}\label{algo:connected}
    \begin{algorithmic}
    \REQUIRE $H$ (a hypergraph)
    \ENSURE components (list of connected components)
    
    \STATE $V \gets graph.vertices\_dict$
    \STATE $E \gets graph.edges\_dict$
    \STATE $n_V \gets |V|$
    \STATE $n_E \gets |E|$
    \STATE $normal\_graph \gets$ Create\_Normal$(n_V + n_E)$ (from \texttt{Hgraph})
    
    \FOR{each $vertex$ in range($n_V$)}
        \FOR{each $edge$ in range($n_E$)}
            \IF{$V[vertex]$ in $E[edge]$}
                \STATE $normal\_graph$.add\_edge$(vertex, edge + n_V)$
            \ENDIF
        \ENDFOR
    \ENDFOR
    
    \STATE $cc \gets normal\_graph$.connected\_components()
    \STATE $components \gets []$ (list)
    
    \FOR{each $cluster$ in $cc$}
        \STATE $component \gets []$ (list)
        \FOR{each $i$ in $cluster$}
            \IF{$i < n_V$}
                \STATE $component$.append($V[i]$)
            \ENDIF
        \ENDFOR
        \STATE $components$.append($component$)
    \ENDFOR
    
    \RETURN $components$
    \end{algorithmic}
\end{algorithm}

\begin{algorithm}[htb!]
\caption{Reduce Hypergraph to Simple Hypergraph}\label{algo:reduce}
    \begin{algorithmic}
    \REQUIRE $H$ (hypergraph)
    \ENSURE $H\_new$ (reduced hypergraph)
    
    \STATE $H\_new \gets H$.copy()
    \STATE $edge\_list \gets$ list($H.edges$)
    
    \FOR{$i$ in range(len($edge\_list$))}
        \FOR{$j$ in range(len($edge\_list$))}
            \IF{$i \neq j$ \AND $edge\_list[j] \subseteq edge\_list[i]$}
                \STATE $edge\_list\_new \gets$ list($H\_new.edges$)
                \IF{$edge\_list[j]$ in $edge\_list\_new$}
                    \STATE $H\_new$.remove\_edge($edge\_list[j]$)
                \ENDIF
            \ENDIF
        \ENDFOR
    \ENDFOR
    
    \RETURN $H\_new$
    \end{algorithmic}
\end{algorithm}

\begin{algorithm}
\caption{Hypergraph Expansion} \label{algo:expand}
    \begin{algorithmic}
        \REQUIRE $H$ (hypergraph), $mode$ (string) 
        \ENSURE $G$ (normal graph)
        \STATE $vertices \gets H.vertices\_dict$
        \STATE $vertices\_swap \gets \{ v: k \textbf{ for } k, v \in vertices.items() \} $
        \STATE $edges = graph.edges\_dict$
        \STATE $vertices\_length = \textrm{len}(vertices)$ 
        \STATE $edges\_length = \textrm{len}(edges)$
        \IF{$mode == \textrm{`clique'}$} 
            \STATE $G = \textrm{ Create\_Normal}(vertices\_length)$
            \FOR{$edge \textrm{ in range}(edges\_length)$}
                \FOR{$i \textrm{ in } Combinations(edges[edge], 2)$} 
                    \STATE $G.add\_edge(vertices\_swap[i[0]], vertices\_swap[i[1]])$ 
                \ENDFOR
            \ENDFOR
        \ELSIF{$mode == \textrm{`star'}$}
        \STATE $G = \textrm{ Create\_Normal}(vertices\_length)$
            \FOR{$edge \textrm{ in range}(edges\_length)$}
                \FOR{$i \textrm{ in } \textrm{Combinations}(edges[edge], 2)$} 
                    \STATE $G.add\_edge(edge + vertices\_length, vertices\_swap[i])$ 
                \ENDFOR
            \ENDFOR
        \ENDIF
    \RETURN $G$
    \end{algorithmic}
\end{algorithm}

\end{document}